\documentclass[10pt,letterpaper]{article}
\usepackage{opex3}
\usepackage{cite}

\begin{document}

\title{Light trapping in a 30-nm organic photovoltaic cell for efficient carrier collection and light absorption}

\author{Cheng-Chia Tsai$^{1}$, Richard R. Grote$^1$, Ashish Banerjee$^1$, Richard M. Osgood Jr.$^{1,2}$, Dirk Englund$^{1,2}$  }

\address{$^1$Department of Electrical Engineering, Columbia University, New York, NY 10027, USA \\$^2$Department of Applied Physics and Applied Mathematics, Columbia University, New York, NY 10027, USA}

\email{ct2443@columbia.edu} 



\begin{abstract}
We describe surface patterning strategies that permit high photon-collection efficiency together with high carrier-collection efficiency in an ultra-thin planar heterojunction organic photovoltaic cell. Optimized designs reach up to $50\%$ photon collection efficiency in a P3HT layer of only 30 nm, representing a 3- to 5-fold improvement over an unpatterned cell of the same thickness. We compare the enhancement of light confinement in the active layer with an ITO top layer for TE and TM polarized light, and demonstrate that the light absorption can increase by a factor of $2$ due to a gap-plasmon mode in the active layer.    
\end{abstract}



\section{Introduction}
Organic Photovoltaics (OPVs) has great potential as a low-cost, lightweight solar cell technology, but a trade-off exists in designing cells that efficiently absorb photons and ones that efficiently collect photogenerated carriers. This trade-off results because of the exciton diffusion length in OPV active layers such as poly(3-hexylthiophene) (P3HT) is typically only 10-15 nanometers while the photon absorption length is at least an order of magnitude larger \cite{nphoton.2009.park, adma.2008.shaw}. Bulk heterojunction devices address the problem through an optically thick active-layer with intercalated or projected electrodes to harvest charge carriers. Another approach, and the one explore here, employs light-trapping techniques to enhance the absorption in an active layer of thickness comparable to the carrier diffusion length \cite{opex.2009.park, opex.2009.timbleston, apl.2008.morfa,apl.2008.lindquist,opex.2011.shen,ol.2007.Panoiu, opex.2011.raman}.

Various techniques have been proposed to increase photon collection such as photonic crystal structures in the active material itself \cite{opex.2009.park, opex.2009.timbleston} or metallic nanostructures developed on the active layer \cite{apl.2008.morfa,apl.2008.lindquist,opex.2011.shen,ol.2007.Panoiu}. Dielectric gratings can confine light in the form of a resonance mode, and enhance photon absorption in the active layer underneath it \cite{opex.2011.raman}. In this letter, we address the light trapping problem by an extensive numerical design study of a patterned top electrode composed of indium tin oxide (ITO). Our results show that the cell geometry can be tuned to simultaneously take advantage of coupling to optical gap modes, the nature of which is discussed in Section 2, as well localized resonances in the ITO rods and photonic crystal modes in the ITO array. By simultaneously optimizing both types of light confinement mechanisms over the relevant spectral ranges, the resulting cell provides strong enhancement of the solar spectrum. In Sections 3 and 4, we find that this gap mode, which arises in an optically thin low-index material sandwiched between two high-index media, can be formed in the OPV absorbing medium surrounded by ITO and a bottom metal contact. Comparing the photon absorption for TE and TM polarized light, we estimate that the gap mode leads to a near-doubling of absorption in the active region. In Section 5, we extend the one-dimensional gratings to two dimensions, thereby eliminating the polarization-dependence of the absorption enhancement. Using a two-dimensional photonic crystal comprised of ITO rods, we find efficient mode conversion from the normal-incident solar radiation into gap modes sandwiched by metallic contacts beneath and the high-index ITO patterned layers on top. The photonic crystal-patterned structure achieves a photon collection efficiency of 40\%, representing a four-fold enhancement above an unpatterned structure with equal absorbing layer thickness. The absorption remains high for angular incidence up to $70^\circ$ from the normal direction. In Section {\it 5.2}, we find that when ITO rods are replaced by cones, the photon collection efficiency can be increased to 50\%. In all instances, a thin continuous layer of ITO remains on the OPV active material to ensure efficient carrier collection. 

\section{Gap-plasmon mode}
A gap-plasmon mode (gap mode) is the enhanced confinement of a guided surface plasmon mode at a dielectric-conductor boundary, achieved by introducing a thin layer of low index dielectric material (a gap) between a high-index dielectric cladding and the conductor, and is due to the refractive index change across the dielectric interface. A similar enhancement of the field confinement has been demonstrated in dielectric slot waveguides \cite{opex.2008.barrios,ieee.2006.mullner,apl.2005.bj}. In this case, a dielectric slab waveguide is modified by inserting a low index dielectric layer -- the slot -- in the middle of the slab; this slot supports a transverse magnetic (TM) mode. If the slot is narrow enough, the profile remains nearly the same as it was in a slab waveguide without the slot, because of the continuity of the electric displacement field,
\begin{equation}
\varepsilon_w E_w=\varepsilon_s E_s, 
\end{equation}
\begin{equation}
\frac{I_s}{I_w}=\frac{|E_s|^2}{|E_w|^2}=\left(\frac{\varepsilon_w}{\varepsilon_s}\right)^2=\left(\frac{n_w}{n_s}\right)^4.
\end{equation}
The subscripts $w$ and $s$ stand for the waveguide and the slot, respectively. The intensity of the normal component of the electric field increases by a factor of $(n_w/n_s)^4$ in the slot, resulting in stronger absorption . 

\begin{figure}[b]
\centering
\includegraphics[width=10cm]{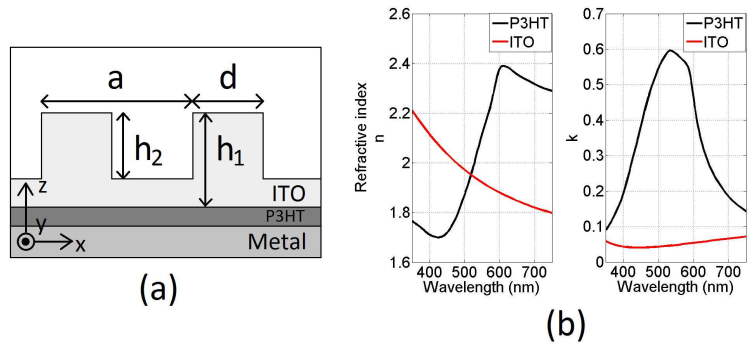}
\caption{(a) Schematic showing the structure composed of a patterning ITO layer on a $30$ nm P3HT absorbing layer. $h_1$ is the thickness of the ITO Layer, $h_2$ is the etching depth of the periodic pattern, $a$ is the period of the pattern, and $d$ is the width of the ridges. (b) The real part ($n$) and the imaginary part ($k$) of the refractive index of P3HT (black) and ITO (red).}
\label{fig:1}
\end{figure}

Similar to the highly confined guided modes in a dielectric waveguide, enhanced standing surface plasmon modes have been observed in a system composed of a narrow low index gap layer sandwiched between a dielectric waveguide with high index and a metal surface on the other side \cite{njp.2008.oulton, natphot.2008.oulton}, and the electromagnetic field can be tightly confined within the gap \cite{opex.2010.avrutsky}. In this letter, we apply the gap-plasmon mode to the dielectric-gap-conductor structure of the OPV device structure proposed in Fig. \ref{fig:1}(a). The periodically patterned dielectric top layer allows for phase matching between incident sunlight and standing surface plasmon gap modes in the P3HT active region for specific wavelengths and angles in regions of the solar spectrum where ITO has a higher index than P3HT. The highly confined gap mode possesses a large in-plane electric field component which leads to a vastly increased optical absorption in the lower index thin film active material. 

\section{1D grating: rectangular ridges}
As illustrated in Fig. \ref{fig:1}(a), the first structure considered consists of a 30 nm active layer placed on a reflective metal layer, and a high-index transparent layer imprinted with periodic patterns on the top of the active layer. The material chosen as the active layer is poly-3(hexylthiophene)(P3HT) and we use a perfect electric conductor (PEC) to model the bottom metal electrode in all our numerical calculations. The real ($n$) and imaginary part ($k$) of the refractive index of the material are shown in the black lines in Fig. \ref{fig:1}(b), Ref. \cite{Ng.TSF.2008}. We consider the solar spectrum (simulated by a blackbody at $5800$ K) from 350 nm to 750 nm, in which P3HT possesses a large absorption coefficient.

To enhance light absorption in the active material, we deposit a transparent layer with high refractive index on the P3HT layer and imprint a one-dimensional pattern of infinitely long square ridges. As defined in Fig. \ref{fig:1}(a), the transparent patterned layer has a thickness of $h_1$, an etch depth of $h_2$, a ridge width of $d$, and a periodicity of $a$. We choose ITO as the material of the top layer in the following discussion to satisfy the previously stated criteria and to act as a transparent top electrode for carrier collection in the OPV. The complex refractive index of ITO is shown in the red lines in Fig.\ref{fig:1}(b) \cite{refractiveindexinfo}. According to the difference between the real index of ITO and P3HT, we can separate the wavelength range into two regimes. In the wavelength range of $\lambda>520$ nm, which we call the {\bf resonance regime,} the enhancement of the absorption results mainly from the resonance modes induced by the ITO photonic crystal structure. As $\lambda<520$ nm, the real part of the index of ITO is larger than that of P3HT so that gap modes can be formed in the P3HT layer and contribute to the absorption enhancement. We call this regime the {\bf resonance-gap hybrid regime} (or, in short, the hybrid regime).

\begin{figure}[b]
\centering
\includegraphics[width=12cm]{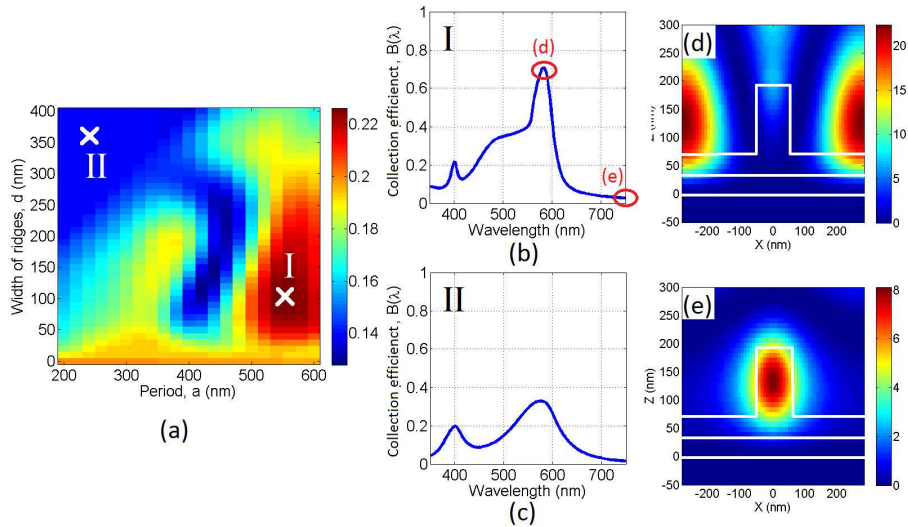}
\caption{(a) Map of the integrated absorbance $A$ versus the period of the 1D pattern, $a$, and the width of the ridges, $d$, for TE polarized light. (b), (c) The photon collection efficiency spectra of the optimal structure and the unpatterned reference. (d), (e) The intensity distribution of the electric field of the optimized case at wavelengths having maximal and minimal collection efficiency.}
\label{fig:2}
\end{figure}

We perform 2D RCWA using the commercially available DiffractMOD simulation package from RSoft Design Group \cite{rsoft} to investigate the dependence of the light absorption in the active layer on all parameters $h_1$, $h_2$, $a$, and $d$. For 2D simulations we use 11 Fourier harmonics in the $x$-direction, and for 3D simulations we use 11 Fourier harmonics in both the $x$- and $y$-direction. For all numerical calculations, plane waves in free space, with wavelengths from $350$ nm to $750$ nm, are used to model the incoming solar light, and periodic boundaries are used in the in-plane ($x-y$) directions. The plane wave source is launched at normal incidence, except for calculations where the angle of incidence is explicitly stated. The figure of merit is defined as 
\begin{equation}
A=\int{d\lambda\left[B(\lambda)\frac{P(\lambda)}{h c/\lambda}\right]}, 
\end{equation}
where $A$ is the total integrated absorbance and $B(\lambda)$ is the collection efficiency of photon, i.e. the calculated fraction of light absorbed in the active layer as a function of wavelength and weighted by the solar irradiance spectrum $P(\lambda)/(hc/\lambda)$, in units of photon flux per unit wavelength. The integral is normalized to unity absorbance.

The absorbance of the planar structure with a $30$ nm P3HT layer and no ITO layer is only $10.13\%$. 

\begin{figure}[b]
\centering
\includegraphics[width=12cm]{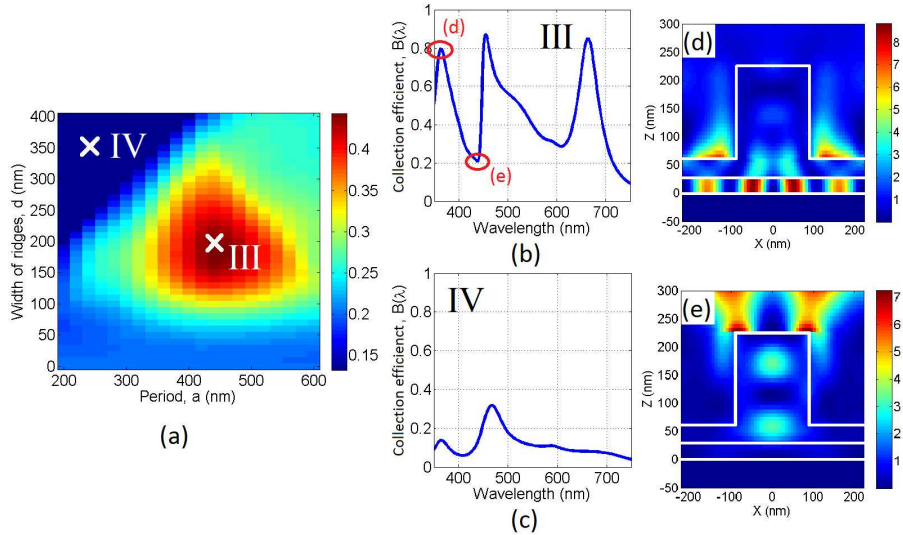}
\caption{(a) Map of the integrated absorbance $A$ versus the period of the 1D pattern, $a$, and the width of the ridges, $d$, for TM polarized light. (b), (c) The photon collection efficiency spectra of the optimal structure and the unpatterned reference. (d), (e) The intensity distribution of the electric field of the optimized case at wavelengths having maximal and minimal collection efficiency.}
\label{fig:3}
\end{figure}

\subsection{TE polarized light}
In our investigation of the photon absorption enhancement by the patterned ITO layers, we first focus on the absorption in the active layer for TE polarized light (with only one $\vec{E}$ component, $E_y$). In this case, neither surface plasmon modes nor gap modes exist. In Fig. \ref{fig:2}(a), we show the overall photon absorbances of the structure with the optimal $h_1=190$ nm and $h_2=150$ nm as determined by the numerical simulations, for different combinations of the ridge period, $a$, and ridge width, $d$. A maximum $A=22.61\%$ occurs at $(a, d)=$(560nm, 100nm) (point I). Compared with a structure with an unpatterned top layer (point II, $A=13.80\%$), the absorbance increases by $64\%$ and its amplitude is more than double that of a structure having only a P3HT layer. The spectra of the photon collection efficiency of the optimized structure and the unpatterned reference are shown in Fig. \ref{fig:2}(b) and \ref{fig:2}(c), respectively. A large peak in the resonance regime with a collection efficiency $B$ of $70\%$ is observed for the optimal case, which agrees with the higher absorbance obtained. Figures \ref{fig:2}(d) and \ref{fig:2}(e) show the intensity distributions of the electric field along the cross-section of the optimal structure at the peak, $\lambda=582$ nm, and the nearest local minimum, $\lambda=750$ nm, respectively. As expected, in both Fig. \ref{fig:2}(d) and \ref{fig:2}(e), resonance modes induced in the ITO grating are observed, but in Fig. \ref{fig:2}(e), there is less intensity distributed in the active layer as the collection efficiency reaches a minimum. This result agrees with the enhancement obtained in a previous study of enhancing light trapping in organic solar cells with dielectric photonic crystal structures as the top-surface layer \cite{opex.2011.raman}. Our approach is to apply this result in combination with a gap mode enhancement, so as to achieve a substantially higher absorption enhancement with a smaller active area. 

\begin{figure}[t]
\centering
\includegraphics[width=6cm]{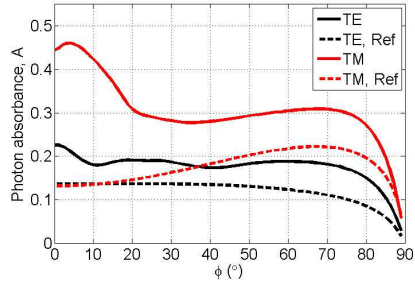}
\caption{Photon absorbance of the rectangular ridge structure with optimized geometry versus incident angle, for TE (black line) and TM polarized light (red line), respectively. The dotted lines show the incident-angle dependence of the absorbance for the unpatterned references}
\label{fig:4}
\end{figure}

\subsection{TM polarized light}
As mentioned in Section 2, because the gap mode results from the enhancement of the surface plasmon mode, we can only induce gap modes with TM polarized light (with only one $\vec{H}$ component, $H_y$). As a result, for comparison, we perform a similar calculation for TM polarized light and show, in Fig. \ref{fig:3}(a), the impact on the photon absorbance of varying the grating period and the ridge width, using the optimized thickness of the ITO layer and etching depth $(h_1, h_2)$=(200 nm, 170 nm). The optimal structure has $(a, d)$=(440 nm, 190 nm) (Point III) and the calculated overall absorbance is up to $44.36\%$, which is a factor of $3.4$ increased from that of the unpatterned reference (Point IV, $A=13.17\%$). Compared with the optimal value for TE polarized light, which does not involve gap modes, the enhancement of absorbance is doubled, which agrees with our expectation of the contribution of the gap modes. The collection efficiency spectrum shows peaks with collection efficiency $B>80\%$ in both the resonance and hybrid regimes for the optimal structure (Fig. \ref{fig:3}(b)) but not for the unpatterned reference (Fig. \ref{fig:3}(c)). Figures \ref{fig:3}(d) and \ref{fig:3}(e) show the intensity of the electric field of the optimal structure for the peak in the hybrid regime at $\lambda=364$ nm and the nearest local minimum at $\lambda=438$ nm. Figure \ref{fig:3}(d) shows the clear presence of a gap mode enhancing the absorption in the active layer; only a small residual field is distributed in the active 
layer in Fig. \ref{fig:3}(e).  

\begin{figure}[t]
\centering
\includegraphics[width=12cm]{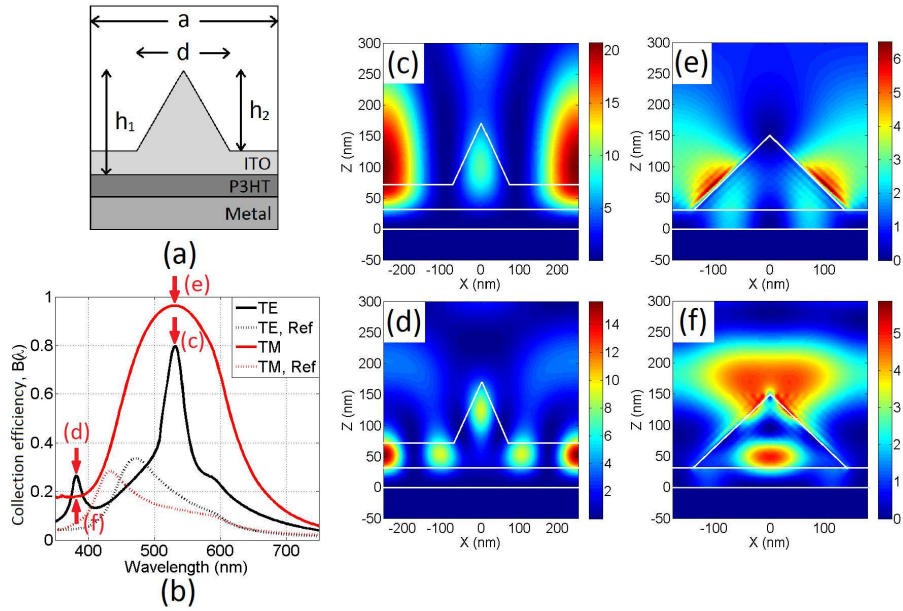}
\caption{(a) Schematic of the ITO grating composed of triangular ridges. (b) The spectral collection efficiencies in the active layer for TE and TM polarized light. (c), (d) The intensities of the electric field for the optimal structure for TE polarized light at $\lambda=$532 nm and 382 nm, respectively. (e), (f) Electric field profiles for TM polarized light at the same wavelengths. }
\label{fig:5}
\end{figure}

\begin{figure}[t]
\centering
\includegraphics[width=6cm]{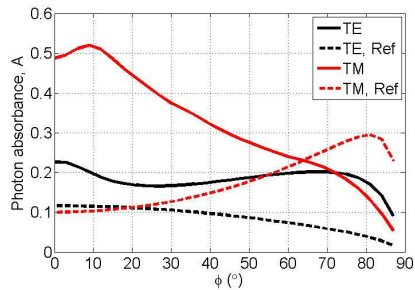}
\caption{Photon absorbance of the triangular ridge structure with optimized geometry versus incident angle, for TE (black line) and TM polarized light (red line), respectively. The dotted lines show the incident-angle dependence of the absorbance for the unpatterned references}
\label{fig:6}
\end{figure}

\subsection{Incident-angle dependence}
One important property of a solar cell is its tolerance to the angle of the incident light. Figure \ref{fig:4} shows the absorbance of the optimized 1D grating cells for TE and TM polarized light, as a function of incident angle. For TE polarized light, the maximal absorbance occurs when light has normal incidence, and for TM polarized light, the absorbance reaches it maximum, $A=46.06\%$, when the light is incident with an angle of $4^\circ$ from the vertical and then drops to its $2/3$ as $\phi>20^\circ$.  However, for both polarizations, the absorbance maintains $40\%$ larger than that of the unpatterned reference until the incident angle is up to $80^\circ$. Note that for TM polarized light, the absorbance of the unpatterned reference increases as the incident angle until $\phi>70^\circ$ because more normal component of electric field can provide stronger gap-plasmon modes in the active layer.

\section{1D grating: triangular ridges}
We replace the square-shaped ridges with a triangular shape to achieve better coupling between the normal-incident electric field and the surface plasmon mode. The structure of the solar cell is shown in Fig. \ref{fig:5}(a). Figure \ref{fig:5}(b) shows the spectral collection efficiency of the optimal structure, which has $h_1=140$ nm, $h_2=100$ nm, $a=520$ nm, and $d=130$ nm for TE polarized light, and $h_1=120$ nm, $h_2=120$ nm, $a=360$ nm, and $d=290$ nm for TM polarized light. The optimal absorbance is $22.64\%$ for TE polarized light, which is enhanced by a factor of 1.95 from the unpatterned reference ($A=11.64\%$). The optimal absorbance for TM polarized light is $48.73\%$, which is $10\%$ higher than the optimal value of the rectangular ridges. A total enhancement of $485\%$ over the unpatterned structure is increased by $43\%$, compared to the $337\%$ enhancement achieved by rectangular-shaped ridges.

\begin{figure}[t]
\centering
\includegraphics[width=13cm]{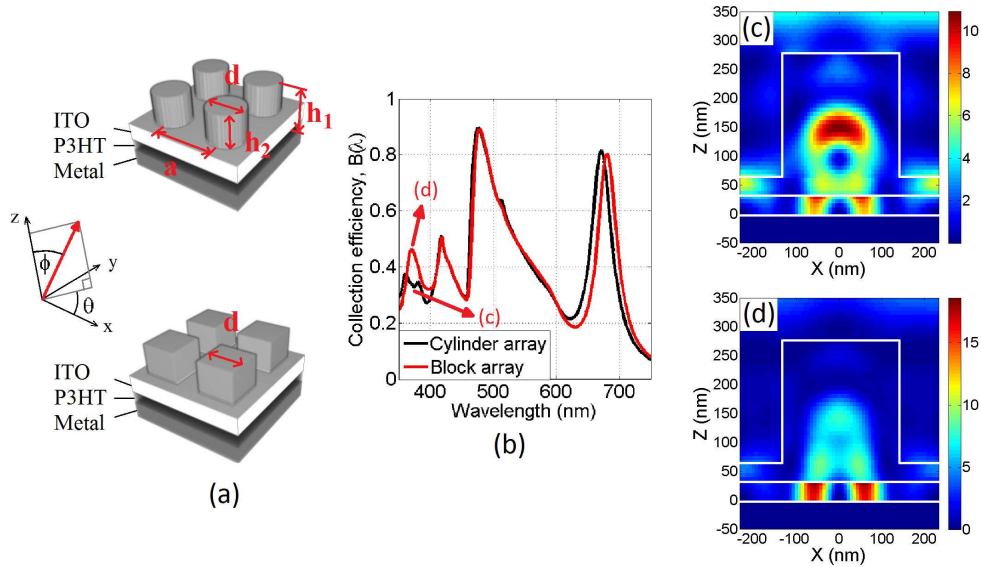}
\caption{(a) Schematic of the cells composed of ITO cylinders or blocks in a square lattice. $h_1$: the thickness of the ITO Layer, $h_2$: the etching depth of the periodic pattern, $a$: the period of the pattern, $d$: the diameter of the rods. (b) The spectral photon collection efficiency for the optimal cylinder and block arrays. (c), (d) The intensity distributions of the electric field of the optimized case, viewed in a cross-section which crosses the axis $y=0$, at $\lambda=370$ nm for the optimal cylinder and block arrays, respectively.}
\label{fig:7}
\end{figure}

Figure \ref{fig:5}(c) and \ref{fig:5}(d) show the intensity distributions of the electric field for peaks at wavelengths $\lambda=$532 nm and 382 nm for TE polarized light. Well-confined resonance modes are observed in the ITO layers. Figure \ref{fig:5}(e) and \ref{fig:5}(f) show the electric field intensity at the same wavelengths for TM polarized light. Beside the resonance modes, surface plasmon modes are observed near the bottom contact and the gap mode is found in the hybrid regime, both of which contribute to the absorption being two times higher than that for TE light.

Figure \ref{fig:6} shows the incident-angle dependence of the photon absorbance of the optimal triangular-ridge gratings. For TE light (the black line),
a maximum exists when the light is normally incident and the absorbance $A$ remains around $0.2$ until the incident angle is larger than $80^\circ$. For TM light (the red light), the photon absorbance reaches its maximum value of $52.03\%$, at $\phi=9^\circ$, and decreases to that of the optimal value for TE light when $\phi<70^\circ$.

\section{2D gratings}
\subsection{Cylinder and block arrays}
To account for the randomly polarized nature of sunlight, we extend the pattern structure from 1D to 2D in order to eliminate the polarization dependence of the absorption enhancement. Figure \ref{fig:7}(a) shows the structure of the 2D patterns that are composed of ITO cylinders and blocks in square lattices. The optimal structure of the cylinder array has $h_1=250$ nm, $h_2=220$ nm, $a=460$ nm, the diameter of the cylinders, $d=260$ nm, and the absorbance is $40.89\%$. The optimal block array has the same $h_1$, $h_2$, and $a$ but a different block width, $d$, of $240$ nm, and the absorbance is $40.55\%$. In both cases, the light absorbance is enhanced by a factor of $\approx 3.4$ over that of an unpatterned ITO layer.

As expected, the cylinder array provides stronger confinement of light than the block array because circular arrays have higher symmetry; however, the absorption enhancement remains almost the same for both structures. We compare the absorption spectra, $B(\lambda)$, of the optimal cylinder and block arrays in Fig. \ref{fig:7}(b). Peaks with high collection efficiency are observed in or near the resonance regime at similar wavelengths for both structures. The peak at $\lambda=670$ nm red shifts for the block array due to the higher effective refractive index. We also observe that a peak exists for the block geometry, but not for the cylinder geometry, in the hybrid regime at $\lambda=370$ nm, where the difference between the index of ITO and P3HT is sufficiently large to support a gap mode. In comparing the intensities of the electric field for the cylinder array at $\lambda=370$ nm in Fig. \ref{fig:7}(c) with that of the block array in Fig. \ref{fig:7}(e), it can been seen that for the cylinder array, the resonance mode of the grating is better confined, but the block array can better couple to gap modes in the active layer because of its less symmetric shape in the in-plane direction, which agrees with the higher collection efficiency obtained.

\begin{figure}[t]
\centering
\includegraphics[width=12cm]{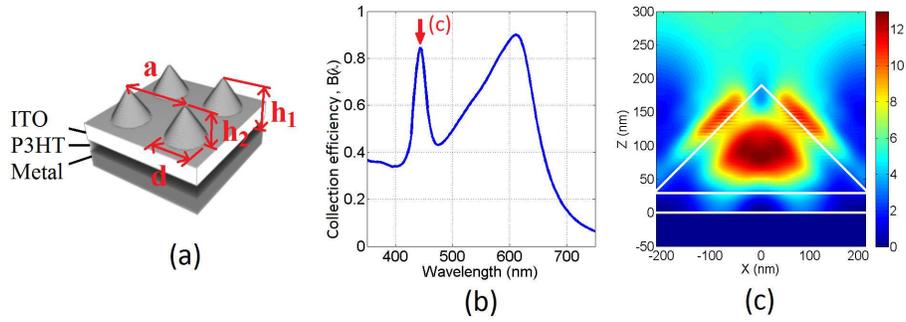}
\caption{(a) Schematic of the cells composed of ITO cones in a square lattice. $h_1$: the thickness of the ITO Layer, $h_2$: the etching depth of the periodic pattern, $a$: the period of the pattern, $d$: the bottom diameter of the cones. (b) The spectral collection efficiency of the optimal structure. (c) The intensity distributions of the electric field of the optimized case, viewed in the plane $y=0$, at $\lambda=444$ nm.}
\label{fig:8}
\end{figure}

\subsection{Cone arrays}
We perform numerical simulations for a 2D grating composed of cones in a square lattice with a similar motivation to the 1D triangular gratings. The optimal structure of the cone array has the dimensions of $h_1=h_2=$160 nm, and $a=d=$430 nm, and the optimal photon absorbance is $49.46\%$, which is enhanced by a factor of $3.7$ from that of the unpatterned reference ($A=13.33\%$). Compared with cylinder or block arrays, the photon absorbance is $25\%$ higher and the enhancement over the unpatterned reference increases by $9\%$. Figure \ref{fig:8}(c) shows the intensity distribution of the electric field for a collection efficiency peak exceeding $80\%$ at $\lambda=444$ nm, which is in the hybrid regime. As can be seen, a gap mode is formed, accompanied with a strongly confined resonance mode in the grating.

\begin{figure}[t]
\centering
\includegraphics[width=11cm]{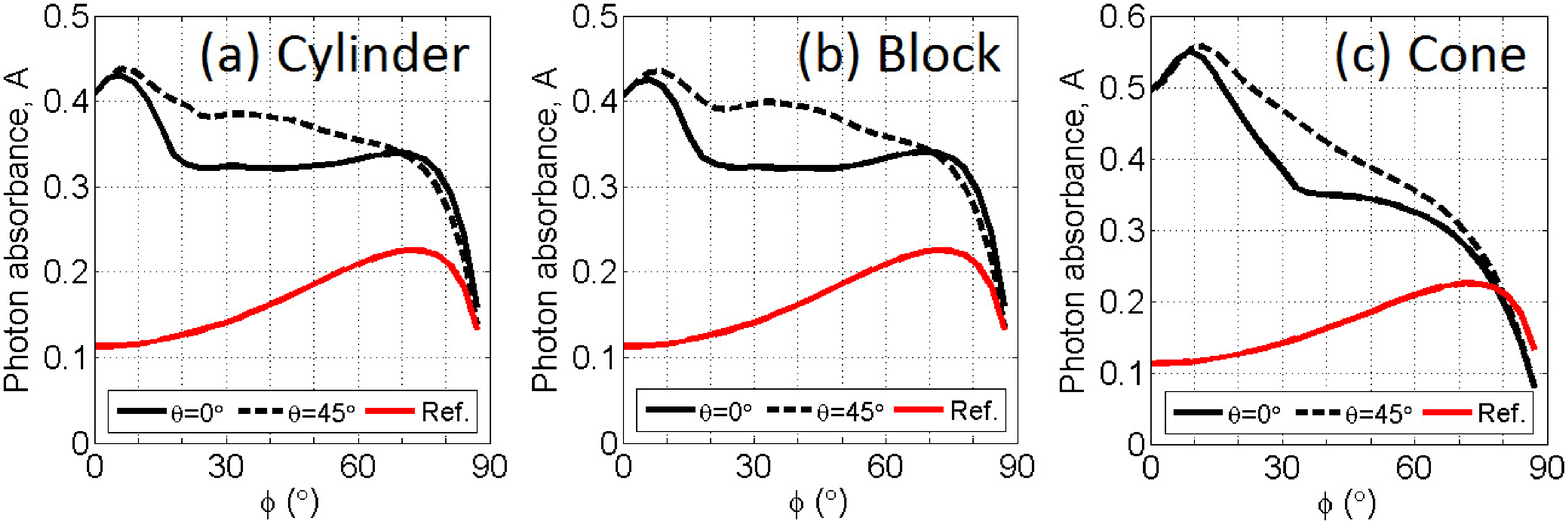}
\caption{Photon absorbance versus incident angle, $\phi$, for the optimal cylinder (a), block (b), and cone arrays (c). The black solid and black dotted lines represent the dependence of $\phi$ along the lattice ($\theta=0^\circ$) and a diagonal ($\theta=45^\circ$), and the red lines is the unpatterned references.}
\label{fig:9}
\end{figure}

\subsection{Incident-angle dependence}
Figures \ref{fig:9} shows the incident angle dependence of the absorbance for the optimal cylinder, block, and cone arrays. The absorbance values of the 2D arrays reach the maxima at $\phi=6-12^\circ$ and are about $5-11\%$ higher than those with normal incident light. The maximal absorbance for the cone array can be up to $A=55.12\%$ when light is incident at an azimuth angle of $9^\circ$ along the lattice, and $A=55.80\%$ at $\phi=12^\circ$ along the diagonal. For all structures, the photon absorbance remains greater than $50\%$ higher than for the unpatterned reference, until $\phi>70^\circ$. Similar to the 1D cases, the absorbance of unpatterned references increase with the incident angle until $\phi>75^\circ$.

\section{Conclusion}
We have extensively studied how different patterning geometries on the top surface of an top-surface ITO contact can effect photon absorption enhancement in an ultra-thin OPV cell. We demonstrate that, in comparison with the enhancement contributed by the resonance modes of a grating only, the enhancement of the photon absorption can be doubled by decreasing the thickness of the active layer to enable gap modes. The photon absorbance can be enhanced by a factor of $3.7$ to $\approx 50\%$ for the optimal 2D cone array, and all types of cells possess large incident-angle tolerance (for angles of $>70^\circ$). This result represents a potential major improvement in optical harvest efficiency for OPVs. We believe this technique can thus lead to a substantial advance in the efficiency of organic solar cells, which has previously been limited by the mismatch between the exciton diffusion length and the absorption length.

\section{Acknowledgments}
This material is based upon work partially supported by the Center for Re-Defining Photovoltaic Efficiency Through Molecule Scale Control, an Energy Frontier Research Center funded by the U.S. Department of Energy, Office of Science, Office of Basic Energy Sciences under Award Number DE-SC0001085. Research was carried out in part at the Center for Functional Nanomaterials, Brookhaven National Laboratory, which is supported by the U.S. Department of Energy, Office of Basic Energy Sciences, under Contract No. DE-AC02-98CH10886. We thank John Kymissis and James Yardley for useful discussions.


\begin{thebibliography}{40}

\bibitem{nphoton.2009.park} S. H. Park, A. Roy, S. Beaupre, S. Cho, N. Coates, J. S. Moon, D. Moses, M. Leclerc, K. Lee, and A. J. Heeger, ``Bulk heterojunction solar cells with internal quantum efficiency approaching $100\%$,'' Nature. Photonics, {\bf 69,} 297--302 (2009).
\bibitem{adma.2008.shaw} P. E. Shaw, A. Ruseckas, and I. D. W. Samuel, ``Exciton Diffusion Measurements in Poly(3-hexylthiophene),'' Adv. Mater. {\bf 20,} 3516--3520 (2008).
\bibitem{opex.2009.park} Y. Park, E. Drouard, O. E. Daif, X. Letartre, P.
Viktorovitch, A. Fave, A Kaminski, M. Lemiti, C. Seassal, ``Absorption enhancement using photonic crystals for silicon thin film solar cells,'' \opex {\bf 17,} 14312--14321 (2009).
\bibitem{opex.2009.timbleston} J. R. Tumbleston, D.-H. Ko, E. T. Samulski and R. Lopez, ``Absorption and quasiguided mode analysis of organic solar cells with photonic crystal photoactive layers,'' \opex {\bf 17,} 7670-7681 (2009).
\bibitem{apl.2008.morfa} A. J. Morfa, K. L. Rowlen, T. H. Reilly, M. J. Romero, and J. van de Lagemaat ``Plasmon-enhanced solar energy conversion in organic bulk heterojunction photovoltaics,'' \apl {\bf 92,} 013504 (2008).
\bibitem{apl.2008.lindquist} N. C. Lindquist, W. A. Luhman, S.-H. Oh, R. J. Holmes, ``Plasmonic nanocavity arrays for enhanced efficiency in organic photovoltaic cells,'' \apl {\bf 93,} 123308 (2008).
\bibitem{opex.2011.shen} H. Shen and B. Maes, ``Combined plasmonic gratings in organic solar cells,'' \opex {\bf 19,} A1201--A1210 (2011).
\bibitem{ol.2007.Panoiu} N. C. Panoiu and R. M. Osgood, Jr., ``Enhanced optical absorption for photovoltaics via excitation of waveguide and plasmon-polariton modes,'' \ol {\bf  32,} 2825--2827 (2007).
\bibitem{opex.2011.raman} A. Raman, Z. Yu, and S. Fan, ``Dielectric nanostructures for broadband light trapping in organic solar cells,'' \opex {\bf 19,} 19015--19026 (2011).
\bibitem{opex.2008.barrios} C. A. Barrios, and M. Lipson, ``Electrically driven silicon resonant light emitting device based on slot waveguide,'' \opex {\bf 13,} 10092--10101 (2005).
\bibitem{ieee.2006.mullner} P. Mullner, and R. Hainberger, ``Structural optimization of silicon-on-insulator slot waveguides,'' IEEE Photon. Technol. Lett. {\bf 18,}, 2557--2559 (2006).
\bibitem{apl.2005.bj} T. Baehr-Jones, M. Hochberg, C. Walker, and A. Scherer, ``High-Q optical resonators in silicon-on-insulatorbased slot waveguides,'' \apl {\bf 86,} 081101 (2005).
\bibitem{njp.2008.oulton} R. F. Oulton, G. Bartal, D. F. P. Pile, and X. Zhang, ``Confinement and propagation characteristics of subwavelength plasmonic modes,'' N. J. Phys. {\bf 10,} 105018 (2008).
\bibitem{natphot.2008.oulton} R. F. Oulton, V. J. Sorger, D. A. Genov, D. F. P. Pile, and X. Zhang, ``A hybrid plasmonic waveguide for subwavelength confinement and long-range propagation,'' Nat. Photonics {\bf 2,} 496--500 (2008).
\bibitem{opex.2010.avrutsky} I. Avrutsky1, R. Soref, and W. Buchwald, ``Sub-wavelength plasmonic modes in a conductor-gap-dielectric system with a nanoscale gap,'' \opex {\bf 18,} 348--363 (2010)
\bibitem{Ng.TSF.2008} A. M. C. Ng, K. Y. Cheung, M. K. Fung, A. B. Djurisic, and W. K. Chan, ``Spectroscopic ellipsometry characterization of polymer¡Vfullerene blend films,'' Thin Solid Film {\bf 517,} 1047--1052 (2008).
\bibitem{refractiveindexinfo} Refractiveindex.Info Refractive Index Database, \url{http://refractiveindex.info/}.
\bibitem{rsoft} RSoft Design Group, Inc., \url{http://www.rsoftdesign.com}.

\end{thebibliography}
\end{document}